\documentclass[%
 ams,
prl,%
 amsmath,amssymb,  
 reprint,%
]{revtex4-1}

\usepackage{graphicx}
\usepackage{dcolumn}
\usepackage{bm}

\begin{document}
\begin{abstract}
Mott insulators with strong spin-orbit coupling have been proposed to host unconventional magnetic states, including the Kitaev quantum spin liquid. The 4$d$ system $\alpha$-RuCl$_3$ has recently come into view as a candidate Kitaev system, with evidence for unusual spin excitations in magnetic scattering experiments. We apply a combination of optical spectroscopy and Raman scattering to study the electronic structure of this material. Our measurements reveal a series of orbital excitations involving localized total angular momentum states of the Ru ion, implying that strong spin-orbit coupling and electron-electron interactions coexist in this material. Analysis of these features allows us to estimate the spin-orbit coupling strength, as well as other parameters describing the local electronic structure, revealing a well-defined hierarchy of energy scales within the Ru $d$ states. By comparing our experimental results with density functional theory calculations, we also clarify the overall features of the optical response. Our results demonstrate that $\alpha$-RuCl$_3$ is an ideal material system to study spin-orbit coupled magnetism on the honeycomb lattice.

\end{abstract}

\title{Spin-orbit excitations and electronic structure of the putative Kitaev magnet $\alpha$-RuCl$_3$}

\author{Luke J. Sandilands}
\altaffiliation{Present Address: Center for Correlated Electron Systems, Institute for Basic Science (IBS) and Department of Physics, Seoul National University, Seoul, 151-742, Republic of Korea\\}

\author{Yao Tian}
\author{Anjan A. Reijnders}
\author{Heung-Sik Kim}
\author{K.W. Plumb}
\affiliation{Department of Physics and Center for Quantum Materials, University of Toronto, Canada M5S 1A7\\}%

\author{Hae-Young Kee}
\affiliation{Canadian Institute for Advanced Research/Quantum Materials Program, Toronto, Ontario MSG 1Z8, Canada}
\affiliation{Department of Physics and Center for Quantum Materials, University of Toronto, Canada M5S 1A7\\}%

\author{Young-June Kim}
\affiliation{Department of Physics and Center for Quantum Materials, University of Toronto, Canada M5S 1A7\\}%

\author{Kenneth S. Burch}
\affiliation{Department of Physics, Boston College, Chestnut Hill, Massachusetts 02467, USA\\}%

\maketitle

\textit{Introduction} -- A variety of novel electronic phases are predicted to emerge in the solid state due to the cooperative action of spin-orbit coupling and electron correlation \cite{doi:10.1146/annurev-conmatphys-020911-125138}. One prominent example is the proposed realization of the Heisenberg-Kitaev model in a strongly spin-orbit coupled Mott insulator on the honeycomb lattice \cite{PhysRevLett.105.027204,PhysRevLett.102.017205}. In this scenario, the combination of spin-orbit coupling and orbital degeneracy leads to the formation of $j_{eff} = 1/2$ pseudospins. The spatial anisotropy inherent to these pseudospins in turn yields bond-dependent, anisotropic exchange interactions that can be mapped onto a generalized Heisenberg-Kitaev model \cite{PhysRevLett.112.077204}, which hosts a variety of unusual magnetic states, including the Kitaev quantum spin liquid \cite{Kitaev20062}. Experimental work in this direction has focused on honeycomb lattice iridates \cite{PhysRevB.82.064412,PhysRevLett.110.076402}, although the electronic structure of these materials is complicated by structural distortions and electron itinerancy \cite{PhysRevB.88.035107}.

In this context, $\alpha$-RuCl$_3$ (hereafter RuCl$_3$) is a promising material for investigating the physics of the spin-orbit coupled Mott insulator on the honeycomb lattice\cite{PhysRevB.90.041112}. This compound crystallizes in a layered structure consisting of planes of edge-sharing RuCl$_6$ octahedra arranged on a honeycomb lattice \cite{Stroganov1957}, although there is some debate on the detailed structure and interlayer stacking \cite{PhysRevB.92.235119}. An important structural detail is that the Ru $d^5$ ion sits in an almost perfect Cl octahedron \cite{Stroganov1957,doi:10.1021/ic00048a025}. The noncubic crystal fields are expected to be small, and the combination of electron-electron interactions and the modest spin-orbit coupling ($\lambda \sim 100$ meV) of the Ru$^{3+}$ ion is thought to be sufficient to induce a Mott insulating, $j_{eff}= 1/2$ ground state \cite{porterfield2013inorganic,PhysRevB.90.041112,PhysRevB.91.241110}. Evidence for a spin-orbit coupled electronic structure is provided by the line shapes and branching ratio observed in x-ray absorption spectroscopy (XAS) \cite{PhysRevB.90.041112}. Recent investigations of the static magnetic properties have pointed towards anisotropic magnetic interactions and therefore to the $j_{eff}=1/2$ state \cite{PhysRevB.91.144420,PhysRevB.91.180401}, while the broad features observed in inelastic magnetic scattering experiments suggests RuCl$_3$ may be close to a Kitaev quantum spin liquid state \cite{PhysRevLett.114.147201,banerjee2015proximate}.

Existing experimental studies of the electronic structure of RuCl$_3$ are, however, limited. A particularly pressing question is whether the $j_{eff}=1/2$ state is well defined in this material. Angle-resolved photoemission measurements found an almost dispersionless feature near the Fermi level which was attributed to weakly dispersing Ru $d$ bands \cite{PhysRevB.53.12769,PhysRevB.50.2095}, while previous optical absorption and reflectivity measurements have identified a series of peaks in the range 0.1 to 10 eV \cite{Guizzetti197934,PSSB:PSSB2220440126,PSSA:PSSA2210710233}. Some controversy exists as to the origin of the optical features and of the magnitude of the fundamental gap, with estimates  ranging from 0.2 to 1 eV \cite{Rojas1983349,PSSB:PSSB2220440126,PhysRevB.90.041112,PhysRevB.53.12769}. However, existing optical studies have considered only a limited temperature range and no electronic Raman scattering data has been reported. More generally, experimental studies of the RuCl$_3$ electronic structure have been hampered by a neglect of spin-orbit coupling (SOC), as well as a lack of electronic structure calculations, which are now available \cite{PhysRevB.90.041112,PhysRevB.91.241110}.

To clarify the low energy electronic structure of this material and determine the relevant energy scales, we have investigated in detail the electronic excitations of RuCl$_3$ using optical spectroscopy and electronic Raman scattering. Importantly, the combination of these methods allows us to obtain a complete picture of the electronic excitations and firmly establish the role of SOC in RuCl$_3$. Optical spectroscopy is well suited to the study of strongly correlated insulators \cite{RevModPhys.83.471,1367-2630-7-1-147} and has provided key insights into the spin-orbit coupled Mott state \cite{PhysRevLett.101.076402,PhysRevLett.101.226402,PhysRevB.88.085125,PhysRevB.89.155115,PhysRevB.80.195110}. Raman scattering is also sensitive to electronic excitations, albeit of different symmetry than optical spectroscopy, and is therefore an excellent complementary probe \cite{Schaack}. The high energy resolution ($<$ 1 meV) of our optical techniques is also advantageous for studying the effects of the modest $\lambda \sim$ 100 meV of $4d$ elements such as Ru. 

Our findings support the notion that Ru is close to the $j_{eff} = 1/2$ limit in RuCl$_3$. The optical and Raman data evince a series of unusual orbital excitations below 1 eV that correspond to transitions involving highly localized and spin-orbit coupled $d$ states of the Ru ion. Analysis of these features allows us to estimate the parameters characterizing the local electronic structure: the spin-orbit coupling ($\lambda$), the effective Hubbard parameter ($U$), and the Hund's coupling ($J_H$). The estimated values are tabulated in Table \ref{params} and demonstrate a well-defined hierarchy of energy scales ($\lambda < J_H <  10Dq$), making RuCl$_3$ an ideal platform for exploring the excitations of the spin-orbit coupled Mott insulating state.  At higher energies, we identify Mott- and charge-transfer-type intersite excitations and find a fundamental optical gap of about 1 eV.   

 \begin{figure}
 \includegraphics[width=\columnwidth]{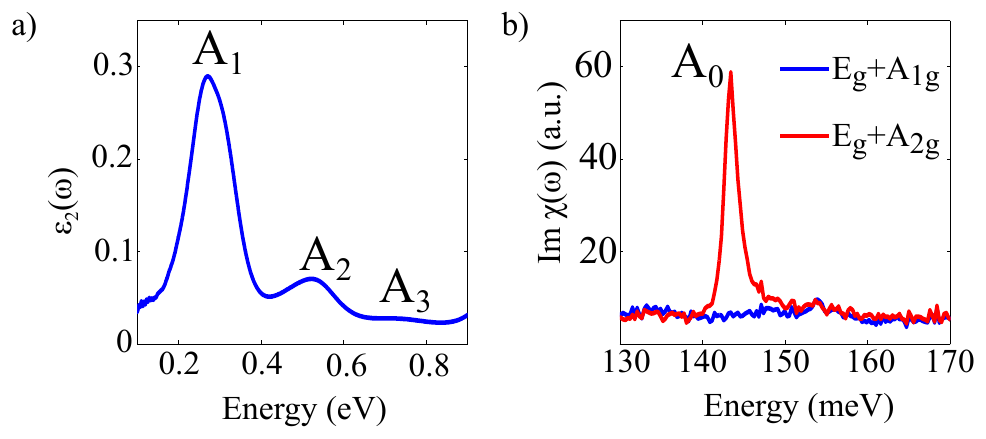}%
 \caption{\label{fig:2}Optical and Raman response of RuCl$_3$. (a) Imaginary part of the dielectric function $\epsilon_2(\omega)$ at 295 K. (b) Raman susceptibility Im $\chi(\omega)$ for two polarization channels at 10 K. Below 1 eV, orbital excitations lead to narrow features, which we label $A_0$-$A_3$, in both $\epsilon_2(\omega)$ and Im $\chi(\omega)$.}
 \end{figure} 

\textit{Experimental details} -- The RuCl$_3$ crystals used in this study were grown by vacuum sublimation of prereacted RuCl$_3$ powder, as described in Ref. \cite{PhysRevB.90.041112}. For the range 0.9 to 6 eV, $\hat{\epsilon}(\omega)$ was determined for an $a$-$b$ crystal face using spectroscopic ellipsometry. For 0.1 to 1.2 eV, we measured the transmittance through a thin ($\sim$ 30 $\mu m$) sample using an FTIR spectrometer described in Ref. \cite{PhysRevB.89.075138}, with light polarized in the crystallographic $a$-$b$ plane. We then extracted $\hat{\epsilon}(\omega)$ using a standard model for the transmittance of a plate sample that accounts for interference effects\cite{kuzmenko:083108}. The $\hat\epsilon(\omega)$ determined in this fashion from the transmittance data is in agreement with the $\hat{\epsilon}(\omega)$ obtained from ellipsometry. Raman scattering measurements were performed in the quasi-back-scattering geometry in both collinear ($E_g + A_{1g}$) and crossed ($E_g + A_{2g}$) polarization geometries using a Raman microscopy system described in references \cite{:/content/aip/journal/apl/98/14/10.1063/1.3573868} and \cite{PhysRevB.86.020403}. Exciting light polarized in the $a$-$b$ plane from a 532 nm laser was focused down to a $\sim$ 10 $\mu$m spot and the power at the sample is estimated to be 500 $\mu$W. A pair of notch filters was used to reject light from the fundamental and the resolution of our Raman spectrometer is estimated to be 0.6 meV. 

\begin{table}\label{params}
\begin{tabular}{ c c c c c}
  \hline
   $U$(eV) & $J_H$ (eV) & 10$Dq$ (eV) & $\lambda$ (eV)  \\
   \hline \hline
   $\geq$ 2.4 &  0.4 & 2.2 & 0.096 \\
\end{tabular}
\caption{\label{params} Parameters characterizing the local electronic structure of the Ru ion. The $10Dq$ value is estimated from the XAS data of reference \cite{PhysRevB.90.041112}. The remaining values follow from the application of the single-ion model to the optical data reported here.}
\end{table}

\textit{Evidence for orbital excitations} -- We first focus on the low energy ($<$ 1 eV) spectra, leaving discussion of the higher energy features, which reflect the band structure, for later. The imaginary part of the dielectric function $\epsilon_2(\omega)$, shown in Fig. \ref{fig:2}(a), consists of three weak and narrow peaks ($A_1$-$A_3$). A low energy excitation ($A_0$) is also visible in the crossed channel of the Raman data shown in Fig. \ref{fig:2}(b). The energy of the four $A$ peaks are large compared to the infrared and Raman active phonon energies ($\sim$ 40 meV) and the Curie-Weiss temperature $(\Theta_{CW}$$\sim$ 3-12 meV) \cite{PhysRevLett.114.147201,J19670001038,PhysRevB.91.180401} of RuCl$_3$, and so the features in our data correspond to electronic excitations. The narrow widths and low intensities (compared to the interband excitations described later) of the $A$ peaks suggests an on-site, orbital origin for these features. Indeed, orbital excitations, also referred to as crystal field excitons \cite{PhysRevB.49.6246} or $dd$ excitations \cite{PhysRevLett.101.157406} in the literature, can often be found below the fundamental absorption edge in transition-metal compounds \cite{1367-2630-7-1-144}. These are local excitations involving the $d$-orbital degrees of freedom of the transition-metal ion. As such, they are formally dipole forbidden, leading to small optical spectral weights, and have energies dictated by the octahedral crystal electric field (10$Dq$), Hund's coupling ($J_H$), and SOC ($\lambda$). As described later, the energies, symmetry, and oscillator strength of these features allow us to assign the $A$ peaks to transitions between spin-orbit coupled multiplets of the Ru ion.

 \begin{figure}
 \includegraphics[width=\columnwidth]{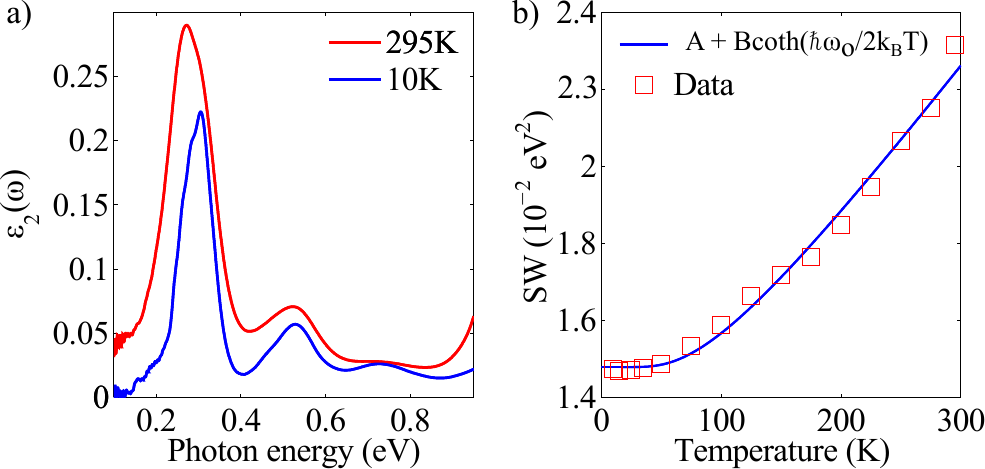}%
 \caption{\label{fig:3} Temperature dependence of the $A_1$-$A_3$ orbital excitations. (a) $\epsilon_2(\omega)$ at 10 and 295 K. (b) Spectral weight $SW = \int_{0.1\ eV}^{0.87\ eV}\omega\epsilon_2(\omega)d\omega$. The smooth decrease in total $SW$ is consistent with phonon-assisted orbital excitations.}%
 \end{figure} 

The Ru site symmetry in RuCl$_3$ is close to $O_h$ and so the electronic states can be classified according to their parity. Due to the dipole selection rule, transitions between even parity $d$ states are not expected to be optically (infrared) active \cite{sugano1970multiplets}. However, such transitions can still acquire a finite optical spectral weight by coupling with an odd parity phonon \cite{PhysRevLett.101.157406}. In other words, the optical absorption process involves a $dd$ excitation together with the simultaneous creation or annihilation of an odd parity phonon. A signature of this mechanism is that the optical response [i.e., $\epsilon_2(\omega)$] should be suppressed at low $T$ due to a reduction in the number of thermally populated phonons. Specifically, the optical spectral weight (SW) acquires a characteristic temperature dependence $ SW  = A + B$coth$ (\hbar\omega_o/2k_BT)$, which reflects the thermal population of the relevant phonon mode(s) with effective energy $\hbar\omega_o$ \cite{PhysRevLett.101.157406}. To verify this, we measured $\epsilon_2(\omega)$ for different temperatures between 10 and 295 K. As shown in Fig. \ref{fig:3} (a), $\epsilon_2(\omega)$ is indeed reduced in going from 295 to 10 K. The integrated $SW = \int_{0.1\ eV}^{0.87\ eV}\omega\epsilon_2(\omega)d\omega$ displayed in Fig. \ref{fig:3} (b) also shows a smooth decrease with temperature, consistent with the phonon-assisted mechanism.  As seen in Fig. \ref{fig:3} (b), the temperature evolution of $SW$ can be well approximated by the expression for phonon-activated absorption with $\hbar\omega_o =$ 22.2$\pm$2.8 meV. This value is in good agreement with a previous study that found infrared active phonons at 21 and 23 meV \cite{J19670001038}. The temperature dependence of $A_1$-$A_3$ is therefore consistent with parity forbidden orbital excitations that become optically active through electron-phonon coupling.

We note that features $A_1$-$A_3$ are significantly broader (width $\sim$ 100 meV) than $A_0$ (width $\sim$ 2 meV). This is because in $\epsilon_2(\omega)$ we observe a $dd$ excitation $plus$ the simultaneous creation/annihilation of an odd-parity phonon. In other words, $\epsilon_2(\omega)$ reflects the (multi-)phonon sidebands of the $dd$ transitions, not the $dd$ transitions themselves \cite{1367-2630-7-1-144}. As a result, the widths of $A_1$-$A_3$ are dictated by the number of phonon sidebands and the relevant phonon energy, rather than the intrinsic lifetime of the underlying $dd$ transition. In contrast, no phonon sidebands are required in the Raman process [i.e., Im $\chi(\omega)$ probes the zero-phonon line] and so $A_0$ is more reflective of the intrinsic line width of the relevant $d$ levels.

\begin{figure}
 \includegraphics[width=\columnwidth]{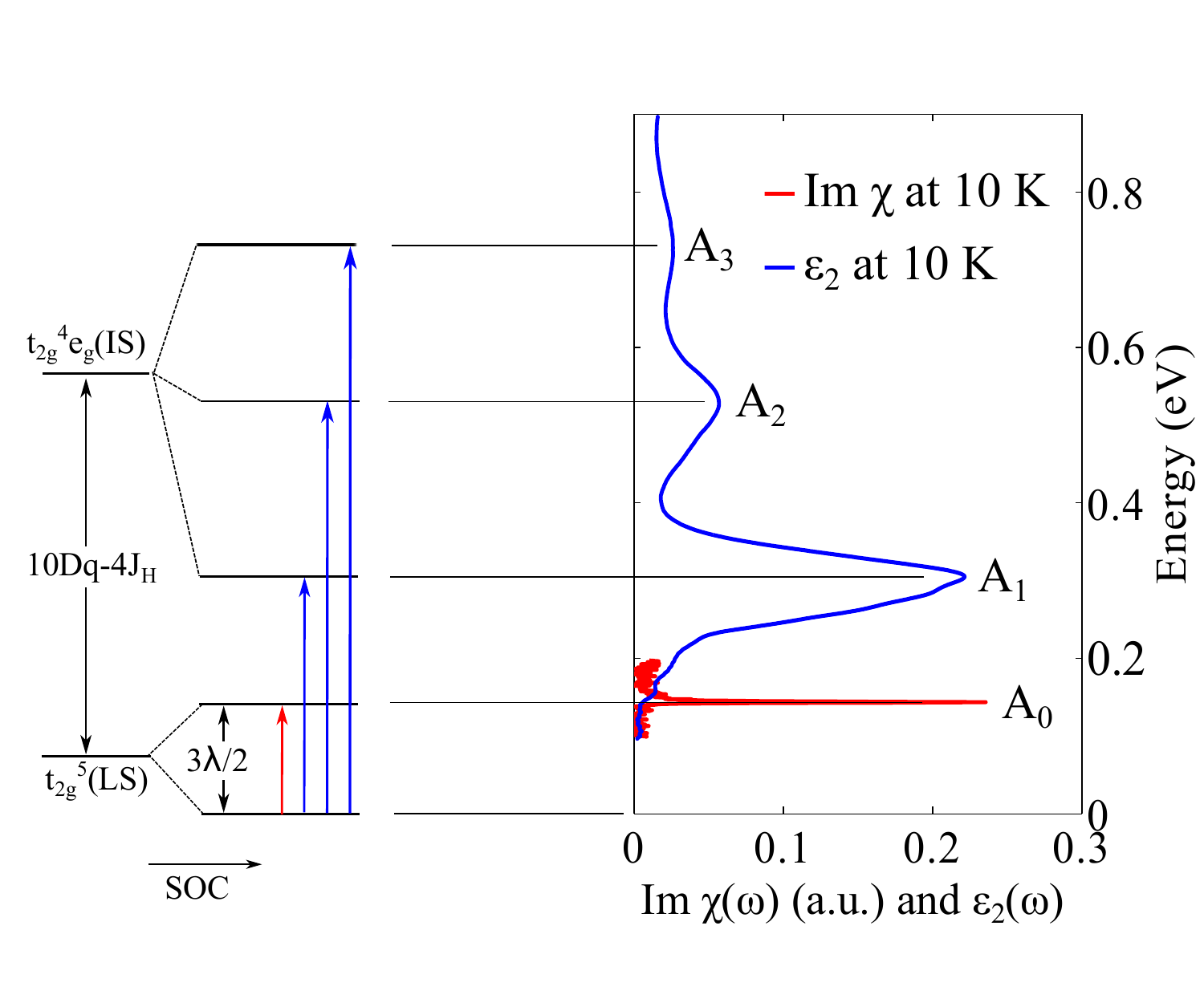}%
 \caption{\label{fig:1}Orbital excitations in RuCl$_3$. In the absence of spin-orbit coupling, the Ru $d^5$ ion is in a low-spin $t_{2g}^5$ ($^2T_2$) configuration. The first excited state is then the intermediate-spin $t_{2g}^4$$e_g$ state. As shown, spin-orbit coupling splits these states, leading to the four transitions observed experimentally.}%
 \end{figure}

\textit{Orbital excitations involving $t_{2g}^5$ configurations} -- To understand the nature of the electronic states involved in the $A$ transitions, we start by considering the low-energy multiplet structure of the Ru$^{3+}$ ion, shown schematically in Fig. \ref{fig:1}. Absent SOC, we expect that the ground state is the low-spin $t_{2g}^5$   ($^2T_2$) configuration, consistent with magnetic susceptibility measurements \cite{PhysRevB.91.144420}. The $^2T_2$ configuration possesses a spin and orbital degeneracy and splits into total momentum $j_{eff}$ states once SOC is introduced. A number of transitions, shown in Fig. \ref{fig:1} as arrows, are therefore expected. For large 10$Dq$ and $J_H$, $^2T_2$  splits into a $j_{eff} = 1/2$ doublet and a $j_{eff} = 3/2$ quartet of energies $-1/2 \lambda$ and $\lambda$, respectively \cite{sugano1970multiplets}. Identifying $A_0$ with the doublet-quartet transition yields $\lambda = 96$ meV, comparable to the Ru$^{3+}$ free ion value \cite{porterfield2013inorganic}. The $A_0$ doublet-quartet transition has in fact been observed in several iridium-based spin-orbit coupled Mott insulators through resonant inelastic x-ray scattering (RIXS) experiments \cite{PhysRevLett.108.177003,PhysRevLett.110.076402,Kim:2014aa} and is usually termed a spin-orbit (SO) exciton. Importantly, the presence of a well-defined SO exciton in RuCl$_3$ places the Ru ion in a $j_{eff}=1/2$ ground state \cite{PhysRevB.89.081109,PhysRevLett.109.167205}.

\textit{Orbital excitations involving $t_{2g}^4$e$_g$ configurations} -- At the energies relevant to the $A_1$-$A_3$ features, the multiplet structure of Ru becomes more complex and excitations involving $e_g$ orbitals become relevant. Specifically, intermediate- and low-spin $t_{2g}^4$$e_g$ configurations are expected near $10Dq-4J_H$ and $10Dq $\cite{PhysRevB.89.081109}. In principle, the high-spin $t_{2g}^3$$e_g^2$ ($^6A_1$) state may be in a similar energy range but is not expected to contribute to $\epsilon_2(\omega)$, as a transition to $t_{2g}^3$$e_g^2$ involves two changes in the single electron occupancies. The energy of these excitations reflects the energy cost for promoting a $t_{2g}$ electron to the $e_g$ levels that is partially offset by the (orbitally averaged) Hund's coupling $J_H$. Following Sham \cite{doi:10.1021/ja00346a028} and using $\lambda = $ 96 meV, the Ru $L_3$ edge XAS data of Ref. \cite{PhysRevB.90.041112} suggests a $10Dq \sim 2.2$ eV. Meanwhile, for Ru $J_H$ is expected to fall in the 0.3 to 0.7 eV range \cite{PhysRevB.90.041112,PhysRevLett.96.256402,PhysRevLett.87.077202,Lee:2006aa}. Intermediate-spin state excitations, in particular $^4T_{1}$ and $^4T_{2}$, are therefore plausibly expected in the region of the experimental $A_1$-$A_3$ states. 

The $^4T_{1}$ and $^4T_{2}$ states retain a spin and orbital degeneracy that should be lifted by SOC. Specifically, they should split into $j_{eff} = 1/2$,  $3/2$, and $5/2$ states. The expected splittings can be estimated from the Land\'e interval rule $E_{j_{eff}}  - E_{j_{eff} -1} = -3\lambda j_{eff} /2$ for $T_{1}$ states \cite{PhysRev.171.466}. For $\lambda = 96$ meV, the estimated splittings between the three states are 216 and 360 meV, comparable to the experimental values of 220 and 200 meV. This assumes that $\lambda$ is the same in both $t_{2g}$ and $e_g$ states. The $A_1$-$A_3$ features can therefore be assigned to the SOC split components of the intermediate-spin $t_{2g}^4$$e_g$ state. Similar SOC-split orbital excitations have also been observed with optical techniques in, for instance, Fe- and Co-based spinels \cite{tokura_spinels}. Within the single-ion model, the lowest-lying component of the intermediate-spin state ($j_{eff} =1/2$) is expected at $E= 10Dq - 4J_H - 15 \lambda/4$ \cite{PhysRev.171.466,PhysRevB.89.081109}. Equating this with the experimental location of the $A_1$ peak of 0.31 eV and taking $10Dq = 2.2$ eV and $\lambda = 0.096$ eV yields a reasonable value of $J_H = 0.38$ eV $\sim 0.4$ eV \footnote{This value is larger than the one used in Ref. \cite{PhysRevB.90.041112}. In this previous work, J$_H$ was simply set to reasonable value as no spectroscopic estimate was available at that time.}. 

The localized, single-ion picture including SOC provides a natural explanation for the number of peaks observed in the optical and Raman data as well as their rough separations. This is shown schematically in Fig. \ref{fig:1}, where the relevant energy levels and transitions are indicated schematically and compared to the experimental data. The parameters derived from the single-ion model are shown in Table \ref{params}. However, we note that the single-ion picture discussed here does not account for electron itinerancy and noncubic crystal fields \cite{PhysRevB.89.081109}. Indeed, the $A_3$ peak is not energetically well separated from the onset of delocalized charge excitations, as we discuss later, and electron itinerancy may therefore be important in determining the energy of this state. Interconfigurational mixing due to noncubic crystal fields or SOC may also be important. The values shown in Table \ref{params} should therefore be interpreted with these considerations in mind. More sophisticated approaches, such as have been applied to iridates and $3d$ transition metal oxides, are required to achieve a more complete and quantitative understanding of the spectrum \cite{PhysRevLett.110.076402,PhysRevLett.109.167205,PhysRevB.89.081109,1367-2630-7-1-144}. 

The comparatively large oscillator strength of $A_1$-$A_3$ in fact provides further evidence for the importance of SOC in RuCl$_3$, even without a detailed understanding of these transitions. Since the Ru site symmetry is close to $O_h$ \cite{Stroganov1957,doi:10.1021/ic00048a025}, transitions to a intermediate-spin $t_{2g}^4$$e_g$ excited state, from which the $A$ peaks in $\epsilon_2(\omega)$ are likely derived, are formally spin and dipole forbidden (i.e., do not conserve spin and involve states of like parity) in the absence of SOC. Spin-forbidden and dipole-forbidden transitions are typically orders of magnitude less intense than their spin-allowed, dipole-forbidden counterparts \cite{sugano1970multiplets}. One might therefore expect that the orbital excitations in RuCl$_3$ should be extremely weak compared to the spin-allowed orbital excitations observed in other materials, such as in $3d$ transition-metal oxides \cite{1367-2630-7-1-144,tokura_spinels}. However, this is not the case, as spin is not a conserved quantity in RuCl$_3$ due to SOC. Our measurements show that the spectral weight $SW = \int_{0.1\ eV}^{0.87\ eV}\omega\epsilon_2(\omega)d\omega$ of the $A_1$-$A_3$ features in RuCl$_3$ is $1.5 \times 10^{-2}$ eV$^2$ at 10 K. As a comparison, the spin-forbidden $^6A_{1g} \rightarrow$ $^4A_{1g}$, $^4E_{g}$ and $^6A_{1g} \rightarrow$ $^4T_{2g}$ transitions in MnF$_2$ have spectral weights in the $10^{-5}$ eV$^2$ range, more than two orders of magnitude lower \cite{PhysRevB.47.15086,Darwish20091470}. The comparatively large $SW$ of the $A_1$-$A_3$ transitions therefore signals a breakdown of the spin selection rule due to the spin-orbit coupled nature of the Ru $d$ states in RuCl$_3$. In fact, the $SW$ of the orbital excitations in RuCl$_3$ is comparable to the values observed for $spin$-$allowed$ transitions in $3d$ transition metal compounds, whose spectral weights typically fall in the $10^{-3}$ to $10^{-2}$  eV$^2$ range \cite{PhysRevLett.101.157406,PhysRevB.84.075160,PhysRevB.54.R11030}\footnote{Refs. \cite{PhysRevB.84.075160} and \cite{PhysRevB.54.R11030} report the optical absorption in units of cm$^{-1}$. To convert to $\epsilon_2(\omega)$ and compare with our data, we assume a refractive index $n$ = 2 in all cases.} 

A last consistency check of our assignment concerns the different orbital character of $A_0$ compared to $A_1$-$A_3$. We note that putative SO exciton $A_0$ only involves  $t_{2g}$ orbitals, while $A_1$-$A_3$ involve both $t_{2g}$ and $e_g$ levels. The $A_1$-$A_3$ excitations are therefore sensitive to the octahedral crystal field and couple strongly to phonons that modify $10Dq$.  As a result, the spectral weight of the $A_1$-$A_3$ features acquired through the electron-phonon mechanism should be large compared to that of $A_0$ \cite{sugano1970multiplets}, in agreement with the data of Fig. \ref{fig:2}. While there is some structure visible near 160 meV [Fig. \ref{fig:2} (a)] that is possibly related to $A_0$, it is weak compared to $A_1$-$A_3$. 

\textit{Comparison with the SO exciton in iridates} -- Despite their common origin, the details of $A_0$ differ in certain respects from the SO exciton observed in iridium-based spin-orbit coupled Mott insulators \cite{PhysRevLett.108.177003,PhysRevLett.110.076402,Kim:2014aa} due to the particular hierarchy of energy scales in RuCl$_3$. In particular, the width of the $A_0$ transition ($\sim$ 2 meV) in RuCl$_3$ is significantly reduced in comparison to, for instance, the 40 meV width of the corresponding excitation in Sr$_2$IrO$_4$ \cite{Kim:2014aa}. This is a consequence of the well-separated energy scales of RuCl$_3$ compared to Sr$_2$IrO$_4$. In the iridates, $\lambda$ is typically larger than the charge gap and so the SO exciton can easily decay into electron-hole pairs. As we shall see later, the onset of the electron-hole continuum is located near 1 eV in RuCl$_3$, significantly higher than $\lambda$. As a result of this separation of energy scales, no electron-hole states are available for the SO exciton to decay into, leading to a correspondingly longer exciton lifetime in RuCl$_3$. This also explains why no orbital excitations have been identified to date in optical studies of existing iridate materials: they are obscured by the more intense, dipole-allowed electron-hole continuum. 

Noncubic crystal fields, such as trigonal distortion, are expected to split the $j_{eff} =3/2$ states into two doublets. As a result, the SO exciton in Li$_2$IrO$_3$ and Na$_2$IrO$_3$ \cite{PhysRevLett.110.076402,PhysRevB.89.081109}, where the trigonal distortion is sizable, shows a characteristic two-peak structure. In contrast, the $A_0$ feature in Fig. \ref{fig:2}(b) displays only a single peak. At first glance, this would imply that the effect of noncubic distortions is small ($\sim$ 1 meV). However, a second possibility is that we do not observe the resulting second peak, either for symmetry reasons (e.g., a selection rule) or that it falls outside the energy range of our experiment. The two components of the SO exciton observed in tetragonally distorted Sr$_2$IrO$_4$ \cite{Kim:2014aa} and in trigonally distorted Na$_2$IrO$_3$ \footnote{H. Gretarsson, private communication} indeed display distinct polarization dependences, with one of the components being enhanced depending on polarization and direction of the incoming x-ray photon. Furthermore, a recent study of (tetragonally distorted) Sr$_2$IrO$_4$ demonstrated that the different $j_{eff} =3/2$ states should contribute to Raman excitations of distinct symmetry \cite{PhysRevB.91.195140}. The fact that we observe a single Raman-active peak ($A_0$) in the crossed channel could be a consequence of this. Addressing the selection rule of the SO exciton would require a detailed understanding of the Raman process, which is beyond the scope of the present study. Thus, while our measurements indicate significant SOC, we are not able to directly quantify the relative strength of any noncubic crystal fields and our value of $\lambda \sim$ 96 meV should be taken with this in mind. 


Another aspect of the iridate RIXS data that bears mentioning is the observation of a sharp peak near the onset of the electron-hole continuum \cite{PhysRevLett.110.076402,Kim:2014aa}. Similar to $A_0$, this feature is narrow (resolution limited in RIXS) and located at $\Gamma$. However, we do not believe these two excitations are related. To our knowledge, two explanations for the RIXS feature have been proposed: an excitonic enhancement of the electron-hole continuum due to long-range Coulomb interactions \cite{PhysRevLett.110.076402} and a mixing between the onsite SO exciton and the continuum of intersite electron-hole excitations \cite{PhysRevB.89.081109}. Both proposals rely on the existence of delocalized charge excitations at similar energies to the sharp feature. However, in RuCl$_3$ delocalized, intersite charge excitations are located at significantly higher energies (above 0.9 eV), as described in the latter part of our manuscript. Further evidence for this point is provided by photoconductivity measurements, which show a strong onset (indicating delocalized states) near 1 eV \cite{PSSB:PSSB2220440126}. We therefore believe that $A_0$ and the sharp RIXS feature do not share a common origin and that $A_0$ instead corresponds to the on-site SO exciton.

 \begin{figure}
 \includegraphics[width=\columnwidth]{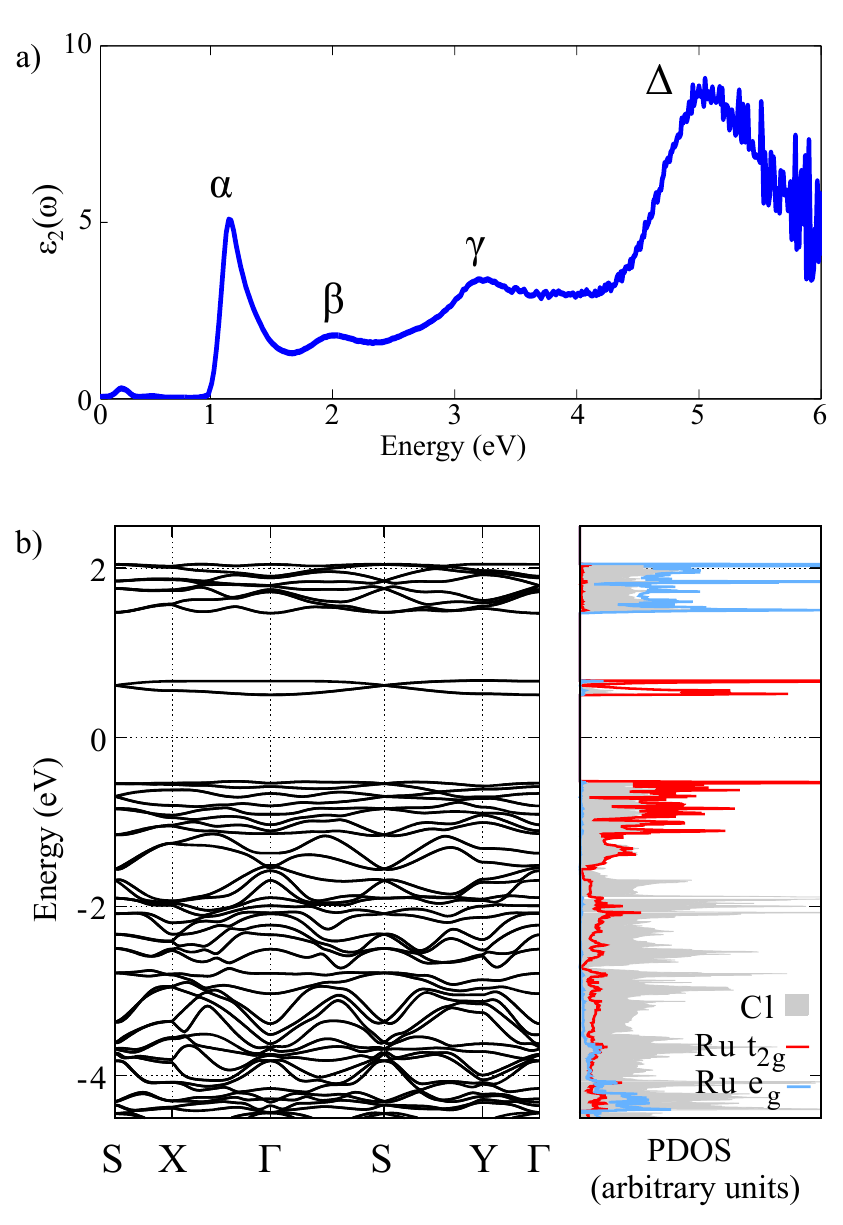}%
 \caption{\label{fig:4} High energy optical response and electronic structure of RuCl$_3$. (a) $\epsilon_2(\omega)$ from 0.1 to 6 eV and (b) density functional theory band structure and orbitally projected density of states (PDOS). }%
 \end{figure} 
 
\textit{Band structure and high energy optical response} -- We now turn away from the orbital excitations and focus on the overall electronic structure, as revealed by the high energy features in $\epsilon_2(\omega)$, shown in Fig. \ref{fig:4}(a). Peaks are visible near 1.16, 2.0, 3.2, and 5.1 eV, which we label $\alpha$, $\beta$, $\gamma$, and $\Delta$, respectively. This portion of our data can be interpreted with the aid of existing photoemission and $ab$ $initio$ studies \cite{PhysRevB.53.12769,PhysRevB.91.241110}. Specifically, photoemission experiments detected three features at binding energies of roughly 0.7, 3.2, and 4.7 eV \cite{PhysRevB.53.12769}. The 0.7 eV feature was identified with narrow bands derived from the Ru $d$ states, while the higher binding energy features were assigned to more dispersive Cl $p$-like bands. This is qualitatively consistent with the LDA+SOC+U density functional theory band structure and orbitally projected density of states (PDOS) displayed in Fig. \ref{fig:4}(b) \footnote{We assumed zigzag magnetic order and $U_{eff} =$ $2.0$ eV. Other details of the calculation can be found in Ref. \cite{PhysRevB.91.241110}}. Near the Fermi level, the occupied states are of primarily of Ru $t_{2g}$ character while Cl $p$-like states are evident at binding energies larger than 2 eV. Importantly, the fact that the lowest-energy feature in photoemission is located at a 0.7 eV binding energy is consistent with the interpretation of the $A_0$-$A_3$ features in terms of orbital excitations, rather than interband transitions. 

Given the 3.2 to 4.7 eV binding energies of the Cl states measured in photoemission, we expect that $d^5 \rightarrow d^6$\underline{$L$} charge transfer excitations should appear at comparable or greater energies in the optical data (\underline{$L$} indicates a ligand hole). The $\alpha$ and $\beta$ peaks at 1.16 and 2.0 eV, which occur at far lower energies, can therefore be interpreted as intersite $dd$ transitions involving neighboring Ru ions (e.g., $d^5 + d^5 \rightarrow d^4 + d^6$). We note that the narrow line shape and asymmetry of $\alpha$ are suggestive of strong electron-hole interaction effects, even if $\alpha$ does not correspond to a true bound state \cite{PhysRevLett.69.1109}. For the low-spin $d^5$ case and neglecting electron-hole interactions effects, the lowest energy intersite excitation occurs at $U-3J_H$ \cite{1367-2630-7-1-147}. Using $J_H \sim 0.4$ eV as determined earlier, we estimate $U = 2.4$ eV. Given that this reasoning neglects electron-hole interactions, $U = 2.4$ eV represents a $lower$ bound. Meanwhile, the intense $\Delta$ peak can be assigned to a charge transfer excitation from Cl to Ru. The origin of $\gamma$ at 3.2 eV is more ambiguous, as both Mott and charge transfer excitations may contribute in this spectral region. This discussion identifies RuCl$_3$ as a Mott-Hubbard, rather than charge-transfer, insulator in the Zaanen-Sawatzky-Allen scheme \cite{PhysRevLett.55.418}. The optical gap, corresponding to the onset of $\alpha$, is about 1 eV. This is larger than reported in some previous works which mistakenly identified the onset of $A_1$ with the true gap \cite{PhysRevB.90.041112,Rojas1983349}.  We caution that a more rigorous theoretical approach, including both band structure and multiplet effects, is needed to confirm this interpretation and obtain a quantitative description of the data \cite{PhysRevB.89.081109,PhysRevLett.109.167205,PhysRevLett.111.246402}.

\textit{Summary}-- We have studied the electronic structure of RuCl$_3$ using optical and Raman scattering spectroscopies. The observed orbital excitations can be understood in terms of well-localized, spin-orbit coupled Ru $d$ states. Our results are broadly suggestive of the importance of spin-orbit coupling in determining the electronic structure of RuCl$_3$ and, by extension, of unconventional magnetic interactions in this material. The well-separated energies of spin-orbit coupling, electron-electron interactions, and charge excitations make RuCl$_3$ an excellent candidate for further theoretical and experimental studies of spin-orbit coupled magnetism on the honeycomb lattice.

\begin{acknowledgements}
KSB acknowledges support from the National Science Foundation (grant DMR-1410846). Research at the University of Toronto was supported by NSERC, CFI, OMRI, and the Canada Research Chair program.
\end{acknowledgements}

%

\end{document}